\documentclass[aps,prl,twocolumn,groupedaddress]{revtex4}
\usepackage{graphicx}
\usepackage{amsmath,amsfonts,amssymb}
\DeclareMathOperator{\erf}{erf}
\usepackage{color}
\begin{document}
\title{Thermal Hawking Broadening and Statistical Entropy
of Black Hole Wave Packet}

\author{Aharon Davidson}
\email{davidson@bgu.ac.il}
\homepage[Homepage: ]{http://www.bgu.ac.il/~davidson}
\author{Ben Yellin}
\email{yellinb@bgu.ac.il}

\affiliation{Physics Department, Ben-Gurion University of
the Negev, Beer-Sheva 84105, Israel}

\date{June 20, 2013}

\begin{abstract}
The quantum mechanical structure of Schwarzschild black
hole is probed, in the mini super spacetime, by means of a
non-singular minimal uncertainty Hartle-Hawking wave packet.
The Compton width of the microstate probability distribution
is translated into a thermal Hawking broadening of the mass
spectrum.
The statistical entropy is analytically calculated using the
Fowler prescription.
While the exact Bekenstein-Hawking entropy is recovered at
the semi classical limit, the accompanying logarithmic tail
gives rise to a Planck size minimal entropy black wave packet.
\end{abstract}

\pacs{}

\maketitle

The Bekenstein-Hawking black hole area entropy \cite{BH}
\begin{equation}
	S_{BH}=\frac{k_B c^3 }{4G\hbar}A
	\label{BH}
\end{equation}
constitutes a triple point in the phase of physical theories,
touching gravity, even beyond general relativity, quantum
mechanics, and statistical mechanics.
However, despite its central role in physics, and several
illuminating derivations \cite{entropy}, its statistical mechanical
roots have not been fully revealed. 
There are only a few exceptions \cite{SV} where one can
actually account, at some limit, for the number of micro states.
As far as the prototype Schwarzschild black hole is concerned,
we still do not know where the micro states are hiding
and even what we are about to enumerate.
A black hole is classically characterized by its event horizon,
but once $\hbar$ is switched on, even an innocent looking
question like 'where is this horizon located' still lacks a
meaningful answer in the quantum or even semi classical sense.
In this paper, we examine the possibility that the Schwarzschild
black hole states are apparently hidden simply because they 
degenerate into one single general relativistic state at the 
$\hbar \rightarrow 0$ limit.
We carry out our analysis within the framework of the mini
super spacetime (a variant of the mini superspace \cite{mini}
limit of Wheeler-DeWitt formalism) without relying on
string theory and/or loop quantum gravity.
We do it by invoking the most classical, that is of minimal
uncertainty, Hartle-Hawking wave packet solution.
The variance of the emerging statistical mass distribution
is then interpreted as thermal Hawking broadening, with the
corresponding semi-classical picture resembling a Gaussian
horizon profile with Compton width (horizon fluctuations were
considered in \cite{Marolf,York}).
The quantum mechanically exposed mass
spectrum is Hawking temperature dependent.
The analytical calculation of the associated statistical entropy 
then calls for the Fowler prescription \cite{Fowler} which has
been designated to deal with temperature dependent energy
levels.

Our starting point is the static spherically symmetric
line element
\begin{equation}
	ds^2=-T(r)dt^2+\frac{dr^2}{R(r)}+S^2(r)d\Omega^2 ~.
\end{equation}
A gauge fixing option, still at our disposal, has to be exercised
with caution, in particular at the mini super spacetime
where the general relativistic action
$\int {\cal R}\sqrt{-g}~d^4 x$ is integrated
out over time and solid angle into the mini action
$\int {\cal L}(T,T^{\prime},S,S^{\prime},R)dr$.
The trail that takes us from here all the way to the reduced
Hamiltonian eq.(\ref{Hred}) has already been presented in the
literature \cite{DY}.
However, for the sake of coherence, the crucial steps must
be briefly outlined.
A word of caution is in order:
Throughout this paper we treat $\int{\cal L}(q,q^{\prime},r)dr$
in full mathematical analogy with $\int{\cal L}(q,\dot{q},t)dt$.
Technically, the $t$-evolution is traded for the $r$-evolution,
both classically as well as quantum mechanically, with the notions
of Lagrangian and Hamiltonian being adapted accordingly.
A similar $t\leftrightarrow r$ technique was adopted by York
and Schmekel \cite{York}. 

Unlike the forbidden gauge prefixing of the 'lapse' function $R(r)$,
which kills the Hamiltonian constraint and introduces an unphysical
degree of freedom, it is apparently harmless \cite{essay} to prefix
(say) $S(r)$ at least at the mini super spacetime level (for an
alternative midi superspace approach, see \cite{Kuchar}).
For example, the gauge choice $S(r)=r$ defines the tenable radial
marker whose geometrical interpretation is $T,R$-independent.
One can verify that up to a total derivative, and up to an overall
factor which can always be absorbed by $T$,
the emerging $r$-dependent mini Lagrangian takes the form
\begin{equation}
	{\cal L}(T,R,R^{\prime},r)=\left(r R^{\prime}
	+R-1\right)\sqrt{\frac{T}{R}}~.
\end{equation}
A simple check verifies that this Lagrangian yields classically the 
Schwarzschild solution and nothing else.
Having in mind the Hamiltonian formalism, however, the trouble
is that the conjugate momenta
$\displaystyle{p_R=\frac{\partial {\cal L}}{\partial R^{\prime}}}$
and $\displaystyle{p_T=\frac{\partial {\cal L}}{\partial T^{\prime}}}$
fail to determine the velocities $R^{\prime}$ and $T^{\prime}$,
giving instead rise to two primary constraints
\begin{equation}
	\phi_1 = p_R-r\sqrt{\frac{T}{R}}\approx 0 ~, \quad
	\phi_2=p_T\approx 0 ~.
	\label{phi12}
\end{equation}
The fact that their Poisson brackets do not vanish, that is
$\displaystyle{\{\phi_1,\phi_2\}=-\frac{r}{2\sqrt{TR}} \neq 0}$,
makes them second class and invites the Dirac procedure
\cite{Dirac} for dealing with constraint systems.
The point is that the naive Hamiltonian
${\cal H}_{naive}=p_R R^{\prime}+p_T T^{\prime}-{\cal L}$
is not uniquely determined, and one may add to it any linear
combination of the $\phi$'s, which are zero, and go over to
${\cal H}^{\star}={\cal H}_{naive}+\sum_i u_i \phi_i $.
Consistency then requires the constraints be constants of
motion, and as such, they must weakly obey
$\displaystyle{\frac{d \phi_i}{dr}\approx 0}$.
Once the $u_i$ coefficients are calculated \cite{DY},
the so-called total Hamiltonian finally makes its appearance
\begin{equation}
	{\cal H}_{total}=(1-R)\left(\sqrt{\frac{T}{R}}
	+\frac{1}{r}\left( p_R -r\sqrt{\frac{T}{R}}\right)
	+\frac{T p_T}{rR } \right)~.	
\end{equation}
Among the associated non-vanishing Dirac brackets, to be
replaced by commutation relations in the quantum theory,
we find the conventional $\{{R,p_R\}}_D=1$, as well as
the unconventional $\{R,T\}_D=2\sqrt{TR}/r$.
Explicitly imposing now the $\phi_{1,2}$ constraints (thereby
importing them to the quantum level), and substituting
\begin{equation}
	T=\frac{p_R Rp_R}{r^2} ~,
	\label{T}
\end{equation}
we are finally led to the reduced on-shell Hamiltonian
\begin{equation}
	\boxed{{\cal H}(R,p,r)=\frac{1}{r}(1-R)p}
	\label{Hred}
\end{equation}
where $p$ stands for $p_R$ for the sake of clarity.
As a preliminary check, one may solve
 $\frac{\partial {\cal H}}{\partial p}=R^{\prime},
~\frac{\partial {\cal H}}{\partial R}=-p^{\prime}$ to confirm 
the Schwarzschild solution $R=1-2m/r,~p=\omega r$ ($\omega$
can be absorbed by rescaling the time coordinate $t$).

The quantization procedure is next.
The $r$-dependent Schr\"{o}dinger equation associated with
the symmetrized reduced Hamiltonian is given by
\begin{equation}
	-i \hbar \frac{1}{2}
	\left((1-R)\frac{\partial }{\partial R}
	+\frac{\partial }{\partial R}(1-R)\right)\psi=
	i \hbar r\frac{\partial }{\partial r}\psi ~.
	\label{Schrodinger}
\end{equation}
Reflecting the fact that the Hamiltonian is linear in the momentum,
our Schr\"{o}dinger equation is apparently $\hbar$-independent.
It is therefore essential to keep in mind the commutation relation
$[R,p]=i \hbar$.
The corresponding 'energy' eigenstates, studied by Berry and Keating
\cite{Berry}
in search for a system where Riemann's $\zeta$-function zeroes
can be physically realized, are not square integrable.
This is however not necessarily a problem here because we are
after the 'most classical'  $\Delta R \Delta p=\frac{1}{2}\hbar$
wave packet solution
\begin{equation}
	\boxed{
	\psi(R,r)=\sqrt{\frac{r}{a}}\left(\frac{2}{\pi}\right)^{1/4}
	e^{\displaystyle{- \frac{r^2}{a^2} (1-R-\frac{b}{r})^2}}}
	\label{psi}
\end{equation}
which is an integral over the Berry-Keating states
\begin{equation}
	\psi(R,r)=\int_{-\infty}^{\infty}
	\frac{f(u) r^{-\frac{i}{\hbar}u}du}{(1-R)^{\frac{1}{2}
	+\frac{i}{\hbar}u}}~,
\end{equation}
for some weight function $f(u)$.
Multiplying $\psi$ by a phase factor
$e^{\displaystyle{{i \omega r (1-R+b/r)}}}$, in charge of
$\langle p \rangle=\hbar \omega r$, is optional, but would not
affect our main conclusions.
The wave packet eq.(\ref{psi}), the first one in a tower of orthonormal
$\Delta R \Delta p=(2n+1)\hbar/2$ wave packets, is non singular, neither at the $r\rightarrow 0$
limit nor at the boundaries where $\psi(\pm \infty,r)\rightarrow 0$.

The general relativity limit is approached when the width
$\displaystyle{\sigma (r)=\frac{a}{2r}}$ of the wave packet tends
to zero.
$\psi^{\dagger}\psi$ becomes in this limit a narrow Dirac delta
function peaked at the classical Schwarzschild solution. 
In turn, recalling that
$\langle T \rangle \sim \langle R \rangle=1-b/r$, we easily
identify
\begin{equation}
	b=\frac{2Gm}{c^2}~.
\end{equation}
The width must then vanish at the $\hbar\rightarrow 0$ limit.
Following Bekenstein's insight, one further expects the width
to be purely quantum mechanical.
It has been argued that adding one bit of information to a
heavy black hole increases its $G$-dependent Schwarzschild
radius by an amount set by its $G$-independent Compton
length scale (strikingly not by Planck scale).
This paves the way for
\begin{equation}
	a=\frac{2\eta\hbar}{mc} ~,
	\label{a}
\end{equation}
where $\eta$ is a dimensionless constant to be determined
below.
The Planck scale enters the game via $ab=4\eta \ell_{Pl}^2$.
For example, associated with the Schwarzschild mass operator
$M=\displaystyle{\frac{c^2 r}{2G}(1-R)}$ are the averages
\begin{equation}
	\langle M\rangle^2=m^2 ~,\quad
	\langle M^2\rangle =m^2+
	\frac{\eta^2 \hbar^2 c^2}{4G^2 m^2}~,
\end{equation}
setting up a fundamental lower bound
$\langle M^2\rangle_{min}=\eta M_{Pl}^2$
which will soon be translated into a minimal entropy.

A close inspection reveals that the black wave packet probability
density $\psi^{\dagger}\psi$ can be directly translated into the
statistical mechanics normalized energy distribution
\begin{equation}
	\boxed{\rho(E,m)=\frac{\sqrt{2}Gm}{\sqrt{\pi}\eta \hbar c^3}
	~e^{\displaystyle{-\frac{2G^2 m^2}{\eta^2 \hbar^2 c^6}
	\left(E-mc^2\right)^2}}}
	\label{rhoE}
\end{equation}
where $E=Mc^2$.
While a positive $m$ is a matter of choice, like in the
Schwarzschild solution, the mass distribution must cover now
the full range $-\infty <M<\infty$, negative masses included.
However, only for $m>0$, the negative masses in the
Gaussian tail come with non-negative probabilities.
This way, the most probable mass is also the average mass, and
the $\langle M \rangle\rightarrow 0$ limit is accessible.

The time is ripe now for the question where is the horizon actually
located?
Counter intuitively, as far as our wave packet is concerned, there
is nothing special going on in the neighborhood of $r= 2Gm/c^2$.
So, quantum mechanically, the answer may well be that there is
no horizon whatsoever; the horizon is just a purely classical
gravitational concept.
Semi classically, however, one may interpret eq.(\ref{rhoE}) as
the quantum mechanical profile of the horizon, with a probability
density $\rho(M,m)$ to find the horizon at radius $2GM/c^2$.
Following this line, it makes sense to define an information extract
function $I(r,m)$.
Classically, no (all) information can be extracted from the black
hole interior (exterior), so $I(r,m)$ must approach the Heaviside
step function $\theta(r-2Gm/c^2)$ at the general relativistic limit.
Upon switching on $\hbar$, $I(r,m)$ is extended to
$I(r,m)=\int_{-\infty}^{\infty} \rho(M,m)\theta(r-2GM/c^2)dM$.
More explicitly
\begin{equation}
	I(r,m)=\frac{1}{2}\left(1+\erf \left(\frac{mc}
	{\sqrt{2}\eta\hbar}
	(r-\frac{2Gm}{c^2})\right)\right)~,
\end{equation}
indicating that partial information can be extracted from
$r<2Gm/c^2$ regions, while some information emanating from
$r>2Gm/c^2$ regions gets blocked.
This may suggest that black hole radiation cannot be purely thermal.
\begin{figure}[h]
	\includegraphics[scale=0.5]{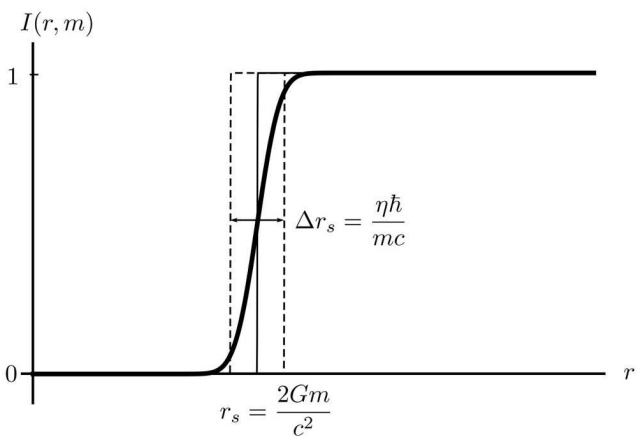}
	\caption{The information extract function $I(r,m)$,
	defined for the semi-classical (large-$m$) horizon
	profile interpretation, replaces the classical Heaviside
	step function $\theta(r-2Gm/c^2)$.}
	\label{Fig1}
\end{figure}

The energy distribution variance
$\Delta E=\sqrt{\langle E^2\rangle-\langle E\rangle^2}$ 
determines the Hawking temperature, to be specific
\begin{equation}
	\Delta E=\frac{\eta \hbar c^3}{2Gm}
	=k_B T_H\equiv \frac{1}{\beta}~,
	\label{kT}
\end{equation}
thereby fixing the parameter
$\displaystyle{\eta=\frac{1}{4\pi}}$, and a posteriori justifying
eq.(\ref{a}).
Eq.(\ref{kT}) defines the thermal Hawking broadening of the black
wave packet.
Such a $\Delta E$ is consistent with Hawking's Euclidean path
integral formalism, where the central role is played by the imaginary
time periodicity
$\displaystyle{c\Delta \tau=\frac{\hbar c}{\Delta E}=
\frac{8\pi Gm}{c^2}}$.
Notice that the entire spectrum eq.(\ref{rhoE}) degenerates into
a Dirac $\delta (E-mc^2)$ function at the GR limit when
$\hbar\rightarrow 0$, thereby hiding all quantum mechanical
degrees of freedom which govern black hole thermodynamics.

Recalling that the average mass $m$ is $T_H$-dependent,
one faces a continuous black hole energy spectrum whose levels are
in fact temperature dependent.
This puts us in a less familiar statistical mechanics territory and calls
for extra caution.
According to Fowler prescription \cite{Fowler}, the naive partition
function should be modified in this case
\begin{equation}
	\sum_n \rho_n e^{-\beta E_n}~\rightarrow~
	\sum_n \rho_n e^{-\beta F_n }~,
	\label{Fowler}
\end{equation}
with the Boltzmann factor being traded for the Gibbs-Helmholtz 
factor.
The Helmholtz free energy function $F$ obeys the Gibbs-Helmholtz
equation
\begin{equation}
	F+\beta\frac{\partial F}{\partial \beta}=E(\beta)
	~\Rightarrow~
	F(\beta)=\frac{1}{\beta}\int_{\beta_0}^{\beta} E(b)db~.
	\label{GHeq}
\end{equation}
$\beta_0$ is a constant of integration.
Whereas a constant energy $E$ returns $\beta F=\beta E+const$
as expected, a temperature dependent energy (say) $E \sim \beta$
remarkably results in
$\beta F \sim \frac{1}{2}\beta^2+const$.
Expressed now in a statistical mechanical language, the semi-classical
face of this coin is familiar from the conventional  (large $m$) approach.
To be specific, the first law $dE=T_H dS$, interpreted as
$dm\sim m^{-1}dS$, implies $S\sim \frac{1}{2}m^2$
rather than prematurely $S\sim m^2$.
In fact, had not we invoked the Fowler prescription eq.(\ref{Fowler}),
we would have wrongly faced, at the large-$m$ regime
$S=2S_{BH}+const$, which is  twice the amount of
the Bekenstein-Hawking area entropy.

We proceed now to calculate the exact quantum mechanical
Schwarzschild black hole entropy.
First, we divide the normalized distribution $\rho (E,m)$ into $N$
equal probability and temperature independent sections, each of which
representing a wide energy level, such that
\begin{equation}
	\int_{E_n}^{E_{n+1}}\rho(E,m)dE=\frac{1}{N}~.
\end{equation}
This equation is formally solved by invoking the inverse error function
$\erf ^{-1}x$, that is
\begin{equation}
	E_n(\beta)=\frac{\hbar c^5 \beta}{8\pi G}
	-\frac{\sqrt{2}}{\beta}\erf ^{-1}(1-\frac{2n}{N})~,
	\label{En}
\end{equation}
for $n=0,1,...,N$.
The condensation of the black hole states at low Hawking temperatures
is now manifest.
A straightforward solution of the differential Gibbs-Helmholtz
eq.(\ref{GHeq}), with $E_n (\beta)$ eq.(\ref{En}) serving as the
source term, reveals the Helmholtz free energy associated with the
$n$-th level
\begin{equation}
	\beta F_n=\frac{\hbar c^5 \left(\beta^2-\beta_0^2\right)}{16\pi G}
	-\sqrt{2}\log{\frac{\beta}{\beta_0}}\erf ^{-1}(1-\frac{2n}{N})~.
\end{equation}

The next step is to calculate the partition function
\begin{equation}
	Z=e^{-\frac{\hbar c^5 (\beta^2-\beta_0^2)}{16\pi G}}
	\sum_{n=0}^N \frac{1}{N} e^{\sqrt{2}\log\frac{\beta}{\beta_0}
	\erf ^{-1}(1-\frac{2n}{N})}~.
\end{equation}
We let $N\rightarrow \infty$, and define a
continuous integration variable $x=\frac{n}{N}$.
The above sum is subsequently replaced by
$\int_0^1 e^{\sqrt{2}\log\frac{\beta}{\beta_0}\erf ^{-1}(1-2x)}dx$,
leading upon integration to
\begin{equation}
	Z=e^{\displaystyle{-\frac{\hbar c^5
	(\beta^2-\beta_0^2)}{16\pi G}
	+\frac{1}{2} (\log\frac{\beta}{\beta_0})^2}} ~.
\end{equation}
The entropy
$\displaystyle{S=k_B (1-\beta \frac{\partial}{\partial \beta})\log Z}$
associated with this partition function is given explicitly by
\begin{equation}
	\boxed{\frac{S(\beta)}{k_B}=\frac{\hbar c^5 
	(\beta^2+\beta_0^2)}{16\pi G}
	+\frac{1}{2}(\log\frac{\beta}{\beta_0})^2
	-\log\frac{\beta}{\beta_0}}
	\label{S}
\end{equation}
where the leading term is identified to be the exact (factor $\frac{1}{4}$
included) Bekenstein-Hawking
area entropy eq.(\ref{BH}).
Note that eq.(\ref{S}) is an exact quantum mechanical formula,
and not just a perturbative expansion.
Whereas, for large-$m$, it only supplements the leading
Bekenstein-Hawking limit by a novel $(\log)^2$ term (various
$\log$-terms have been discussed in the literature \cite{log}),
it opens a new window into small-$m$ black hole thermodynamics.
\begin{figure}[h]
	\includegraphics[scale=0.65]{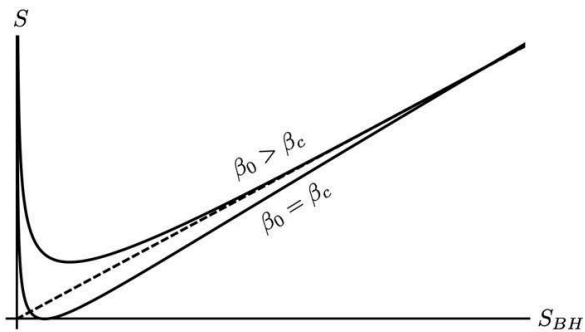}
	\caption{The entropy $S$ matches Bekenstein-Hawking
	$S_{BH}$ (straight dashed line) at the semi classical regime.
	A minimal entropy, $S_{min}=0$ for $\beta_0=\beta_c$,
	connects negative and positive specific heat branches,
	establishing an UV/IR connection.
	If $\beta_0<\beta_c$ (not plotted), these branches get
	disconnected.}
	\label{Fig2}
\end{figure}

The parameter $\beta_0$, which marks the inverse temperature
where the Helmholtz free energy vanishes by construction, is still
arbitrary at this stage.
While $S(\beta)$ is bounded from below, its
minimum $S_{min}$ acquires, via the $(\log)^2$-term, a non-trivial
$\beta_0$ dependence.
At this stage, we can give no general rule for fixing $\beta_0$.
The ambiguity has its counterpart in the use of the Gibbs-Helmholtz
equation to derive the free-energy from the true energy.
To be practical, however, we need to know, for instance, the black
hole entropy at some one particular Hawking temperature.
Having this in mind, we notice that there exists a critical value for
$\beta_{0}$, numerically calculated to be
\begin{equation}
	\beta_{c}^2=s \frac{4\pi G}{\hbar c^5} ~,
	\label{betac}
\end{equation}
$s \simeq 0.404$ obeying
$\log(s^{-1}(\sqrt{5-2s}-1))=3-\sqrt{5-2s}$,
for which the minimal entropy vanishes
\begin{equation}
	S_{min}=0 ~.
	\label{Smin}
\end{equation}
It describes what one can refer to as a non-degenerate ground
state \cite {BM} Schwarzschild black hole.
Associated with its Planck scale average mass
$m_{min}=\frac{\sqrt{s}}{4\sqrt{\pi}}M_{Pl}$
are mass fluctuations of the same order of magnitude.

$\beta_0>\beta_c$ implies $S_{min}>0$.
While the $\beta>\beta_{min}$ branch exhibits a familiar
black hole feature, namely a negative specific heat
$C=-\beta \frac{\partial S}{\partial \beta}<0$, it gets
smoothly connected with a novel branch, associated with
the $\beta<\beta_{min}$ regime, for which the specific heat
is counter intuitively, at least in the black hole sense, positive.
The emerging, what seems to be a realization of the so-called
UV/IR connection \cite{UVIR}, takes us into an intriguing yet
unfamiliar territory.
If $\beta_0<\beta_c$, on the other hand, one encounters a 
region characterized by a negative entropy.
We find this situation unacceptable recalling the fact that the
entropy,  being a logarithm measure of the total number
of configurations, is non-negative definite.
In other words, there is a gap disconnecting now the two
branches mentioned earlier.
Based on the above arguments, $\beta_0=\beta_c$ becomes
our choice of preference if the ground state is non-degenerate.


To summarize, the Schwarzschild black hole states are
apparently hidden simply because they degenerate into one
single general relativistic state at the $\hbar \rightarrow 0$
limit.
Once $\hbar$ is switched on, the Compton width of the
Hartle-Hawking wave packet gets revealed, with Schwarzschild
geometry becoming just the most probable (as well as the average)
solution.
Converting the probability density into a statistical mechanics
energy distribution, the variance of the latter means thermal
Hawking broadening of the wave packet, thereby paving the
way for calculating the statistical entropy (adopting Fowler's
prescription on technical grounds).
While the exact Bekenstein-Hawking entropy is consistently
recovered at the semi classical limit, its logarithmic tail gives
rise to a minimal entropy Planck size black wave packet.
The inclusion of a cosmological constant $\Lambda$ is
straightforward, including the BTZ and AdS cases, but several
other questions are still open and deserve further clarification.
For example, the role played by the negative masses in the
spectrum, the physical meaning of the $0\leq m<M_{Pl}$ sub
Planckian branch (unphysical? quantum foam? elementary
particles?), and the possibility of elevating the minimal entropy
configuration to the level of the fundamental black hole building
block.
It remains to be seen if our work is somehow relevant
for artificial black holes \cite{artificial} as well.

\acknowledgments
{Special thanks to BGU president Prof. Rivka Carmi for her
kind support.
Valuable conversations with Prof. Doron Cohen, Dr.
Ilya Gurwich and Dr. Shimon Rubin are appreciated.}

\end{document}